# Preliminary Demonstration of Diamond-GaN *pn* Diodes via Grafting

Jie Zhou[1,8,a)], Yi Lu[1,a)], Chenyu Wang[2,a)], Luke Suter[3,a)], Aaron Hardy[3,a)], Tien Khee Ng[5,a)], Kai Sun[4 a)], Yifu Guo[6], Yang Liu[1], Tsung-Han Tsai[1], Xuanyu Zhou[2], Connor S Bailey[7], Michael Eller[7], Stephanie Liu[7], Zetian Mi[6], Boon S. Ooi[5,9], Matthias Muehle[3], Katherine Fountaine[7], Vincent Gambin[7,b)], Jung-Hun Seo[2,b)], and Zhenqiang Ma[1,b)]

[1]*Department of Electrical and Computer Engineering, University of Wisconsin-Madison, Madison, Wisconsin, 53706, USA*

[2]*Department of Materials Design and Innovation, University at Buffalo, The State University of New York, Buffalo, NY 14260, USA*

[3]*Fraunhofer USA Inc, Center Midwest East Lansing, MI 48824, USA*

[4]*Department of Materials Science and Engineering, University of Michigan, Ann Arbor, MI 48109, USA*

[5]*Department of Electrical and Computer Engineering, King Abdullah University of Science and Technology, Thuwal 23955-6900, Saudi Arabia*

[6]*Department of Electrical Engineering and Computer Science, University of Michigan, Ann Arbor, Michigan 48109, USA*

[7]*Northrop Grumman Corporation, Redondo Beach, CA 90278, USA*

[8]*Department of Electrical and Computer Engineering, The University of Texas at Dallas, Richardson, TX 75080, USA*

[9]*Department of Electrical, Computer, and Systems Engineering, Rensselaer Polytechnic Institute, Troy, NY 12180, USA*

a) These authors contributed equally to this work.
b) Author to whom correspondence should be addressed. Electronic mail: mazq@engr.wisc.edu or vincent.gambin@ngc.com, or katherine.fountaine@ngc.com or junghuns@buffalo.edu

# ABSTRACT

Ultrawide bandgap (UWBG) semiconductors exhibit exceptional electrical and thermal properties, offering strong potential for high-power and high-frequency electronics. However, efficient doping in UWBG materials is typically limited to either *n*-type or *p*-type, constraining their application to unipolar devices. The realization of *pn* junctions through heterogeneous integration of complementary UWBG (or WBG) semiconductors is hindered by lattice mismatch and thermal expansion differences. Here, we report the preliminary demonstration of diamond–GaN heterojunction *p–n* diodes fabricated via *grafting*. A single-crystalline $p^+$ diamond nanomembrane was integrated onto an epitaxially grown *c*-plane $n^-/n^+$ GaN substrate with an ultrathin ALD–$Al_2O_3$ interlayer. The resulting diodes exhibit an ideality factor of 1.55 and a rectification ratio of ~$10^4$. Structural and interfacial properties were examined by AFM, XRD, Raman, and STEM, providing critical insights to guide further optimization of diamond–GaN *pn* heterojunction devices.

Ultrawide bandgap (UWBG) materials, including oxides (e.g., $Ga_2O_3$), nitrides (e.g., Al(Ga)N, *c*-BN), and diamond ($C^{sp3}$, distinguished from graphene, $C^{sp2}$)[1–3], exhibit high critical electric fields, enabling superior Baliga and Johnson figures of merit for high-power and high-frequency applications. Diamond combines high carrier mobility with exceptional thermal conductivity, offering substantial improvements in power handling, switching speed, and thermal management. Despite these advantages, effective doping is typically limited to either *p*-type (e.g., $C^{sp3}$) or *n*-type (e.g., $Ga_2O_3$, AlN), restricting UWBG materials to unipolar applications. Actively integrating *p*-type and *n*-type UWBG materials is, therefore, critical to unlock the full potential of the UWBG family with the versatility typically only associated with narrow bandgap semiconductors.

Heterogeneous integration, however, often faces major challenges. i) High-quality heterointerfaces: conventional heteroepitaxy requires close lattice matching[4–6], whereas direct wafer bonding or fusion[7,8] often introduces dangling bonds, lattice distortions, and high defect densities. ii) The availability of single-crystalline transferable UWBG materials: although bulk wafer bonding has been demonstrated, it involves significant material loss and processing difficulty due to the rigidity and chemical inertness of UWBG crystals. In contrast to mature semiconductors such as Si, Ge, and GaAs, which can be readily released as thin films or membranes for heterogeneous integration, scalable routes to obtain transferable UWBG semiconductors remain limited[9–13]. It is noted that, while diamond has been integrated with GaN and other UWBG materials, it typically serves only as a mechanical support or heat spreader[14–17], rather than as an electrically active component. Deposition methods like sputtering[18–21] or chemical vapor deposition (CVD)[22] have enabled $Ga_2O_3$-based heterojunctions but are not broadly applicable to other UWBG or even WBG systems, owing to severe lattice and thermal expansion mismatches.

Semiconductor grafting[23–36] offers a promising route for forming high-quality, lattice-mismatched heterojunctions across narrow bandgap, WBG, and UWBG materials. The key is double-sided interface passivation, typically via an ultrathin dielectric interlayer deposited by ALD[28,32], or formed through surface chemical oxidation[29,36] or interface nitrigenation[37]. This layer suppresses trap states and interdiffusion, accommodates differences in thermal expansion and mechanical properties[23], and permits efficient electron and hole tunneling due to its minimal thickness.

Here, we extend grafting to realize an electrically active, single-crystalline $C^{sp3}$–GaN heterojunction. We employ Raman spectroscopy, X-ray diffraction (XRD), and scanning transmission electron microscopy (STEM) to examine the heterojunction, confirming high crystallinity, structural integrity, and uniform formation of an ultrathin ALD–$Al_2O_3$ interlayer. Circular *pn* diodes fabricated from this heterojunction exhibit an ideality factor of 1.55, a rectification ratio of ~$10^4$, a turn-on voltage of ~4 V, and leakage current density as low as $10^{-7}$ A·$cm^{-2}$. The analysis of the heterojunction and diode behavior provides insights for further optimization of this class of heterojunction devices.

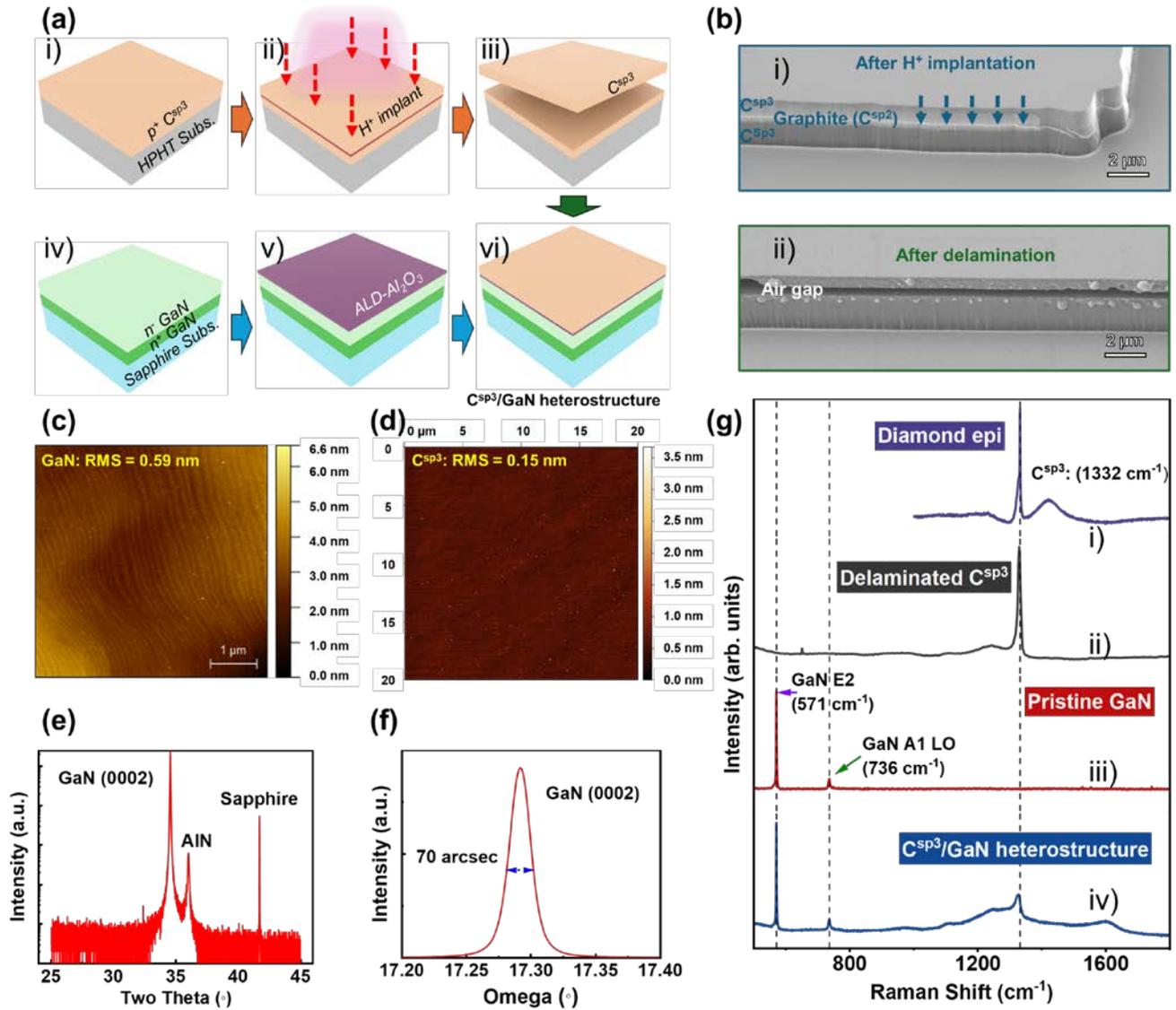

**Figure 1. Characterization of single-crystalline $C^{sp3}$/GaN heterojunction.** (a) Schematic of the fabrication process. (b) SEM images of the diamond membrane: (i) after hydrogen ion implantation and (ii) following high-temperature annealing, showing NM delamination. (c, d) AFM images of GaN and diamond surface morphology. (e, f) XRD 2θ scan and rocking curve (ω-scan) of the Ga-face *c*-plane GaN substrate. (g) Raman spectra of (i) pristine $p^+$-diamond, (ii) delaminated diamond NM, (iii) pristine GaN, and (iv) grafted $C^{sp3}$/GaN heterojunction.

**Figure 1(a)** schematically illustrates the fabrication of single-crystalline diamond (SCD, hereafter referred to as diamond, $C^{sp3}$) nanomembranes (NMs) and their integration into the grafted $C^{sp3}$/GaN heterojunction. A $p^+$-type diamond layer with a thickness of 5.7 μm was epitaxially grown on a nominally (100)-oriented, intrinsic diamond substrate (7 × 7 mm$^2$), and exhibited a sheet resistance of ~19.0 Ω/□ and bulk resistivity of $1.08 \times 10^{-2}$ Ω·cm [**Fig. 1(a i)**], as measured by room temperature four-point probe. The 1-hour $p^+$ growth process was carried out using microwave plasma chemical vapor deposition (MPCVD) with growth conditions described elsewhere[38]. Post-growth chemical–mechanical polishing of the $p^+$ epilayer was carried out using a custom polishing tool described elsewhere[39], with an updated slurry composition that contains a Fenton's reagent as the oxidant, and $SiO_2$ abrasive particles, reducing the surface roughness to the sub-nanometer scale [**Fig. 1(d)**]. Atomic force microscopy (AFM) revealed a root-mean-square (RMS) roughness of 0.15 nm over a 20 × 20 μm$^2$ area.

To release the diamond NM from the substrate, hydrogen ion implantation was performed under 225 keV with a dose of $2\times10^{17}$ cm$^{-2}$ at room temperature, optimized using Athena® simulations (Supplementary **Fig. S1**), at Michigan Ion Beam Laboratory. Implantation creates an approximately 50 nm thick- high concentrated hydrogen layer that is located 1.5 μm below the diamond surface. Cyclic high temperature annealing with water vapor and $N_2$ was performed in a furnace at 600 °C for 24 hours under atmospheric pressure to promote a chemical reaction that forms hydrogen bubbles, leaving a physical gap between the diamond NM and the substrate while maintaining single crystallinity in both [**Fig. 1(a ii)**]. The separated diamond NM [**Fig. 1(a iii)**, Supplementary **Fig. S2**] was subsequently transferred onto the GaN epilayer [**Fig. S9**]. Scanning electron microscopy (SEM) images of the ion-implanted and delaminated SCD are shown in **Figs. 1(b i)** and **1(b ii)**, respectively.

The $n^+/n^-$ GaN (250 nm/1.55 μm) host substrate [**Fig. 1(a iv)**], grown by standard metalorganic chemical vapor deposition (MOCVD) above 700 °C, exhibited excellent surface quality. AFM revealed an RMS roughness of 0.59 nm over a 5 × 5 μm$^2$ area [**Fig. 1(c)**], ensuring intimate contact with the diamond NM during transfer printing. High-resolution XRD confirmed the high crystallinity of the GaN epilayers, with the (0002) peak at 2θ = 34.584° and a FWHM of 0.035° [**Fig. 1(e)**], and a rocking curve width of 70 arcsec [**Fig. 1(f)**].

Prior to integration, the GaN substrate underwent RCA cleaning to remove contaminants and native oxides, followed by deposition of an $Al_2O_3$ interlayer via atomic layer deposition (ALD) in an $N_2$-filled glovebox (Ultratech/Cambridge Nanotech Savannah S200). Five ALD cycles, calibrated at ~0.1 nm/cycle, yielded the ultrathin dielectric [**Fig. 1(a v)**]. Further details on GaN epitaxy, surface preparation, and ALD parameters are provided in our prior reports[40–43], which used the same GaN epi as in this study. The delaminated diamond NM was then picked up with a polydimethylsiloxane (PDMS) stamp, transfer-printed onto the $Al_2O_3$-coated GaN [**Fig. 1(a vi)**], and vacuum annealed at 350 °C for 3 h to form chemical bonds.

Raman spectroscopy tracked the material evolution at key fabrication stages [**Fig. 1(g)**]. The pristine $p^+$-diamond epi showed the characteristic first-order diamond line at 1332 cm$^{-1}$ and a minor transpolyacetylene peak near 1428 cm$^{-1}$ [**Fig. 1(g i)**], originating from the graphite layer induced by hydrogen implantation at the bottom side of the diamond NM. This peak diminished after cyclic annealing [**Fig. 1(g ii)**], indicating complete removal, via oxidation, of the buried graphite layer[44,45]. The pristine GaN substrate exhibited $E_2$ (571 cm$^{-1}$) and $A_1$(LO) (736 cm$^{-1}$) modes [**Fig. 1(g iii)**], confirming its single crystallinity. In the grafted $C^{sp3}$/GaN heterojunction, both GaN and diamond signatures were preserved [**Fig. 1(g iv)**], demonstrating that material integrity was maintained after transfer and bonding.

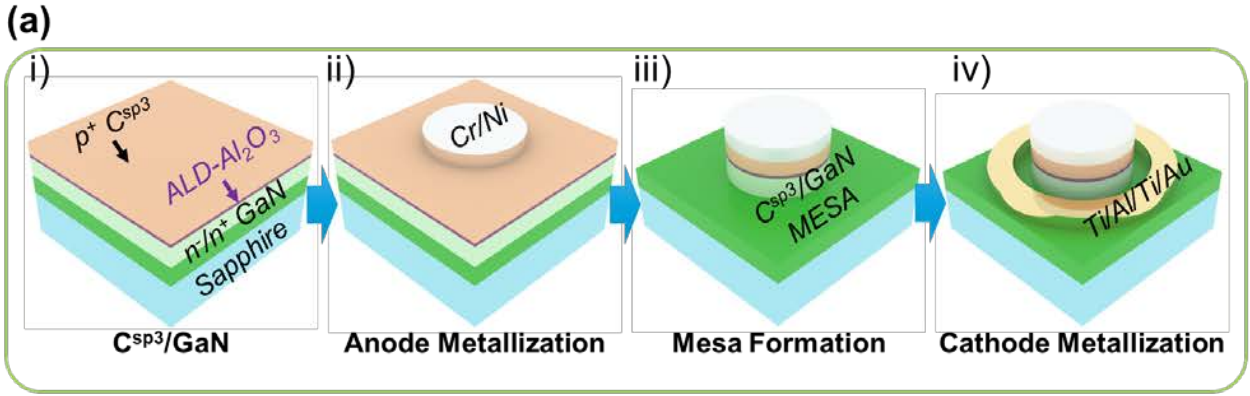

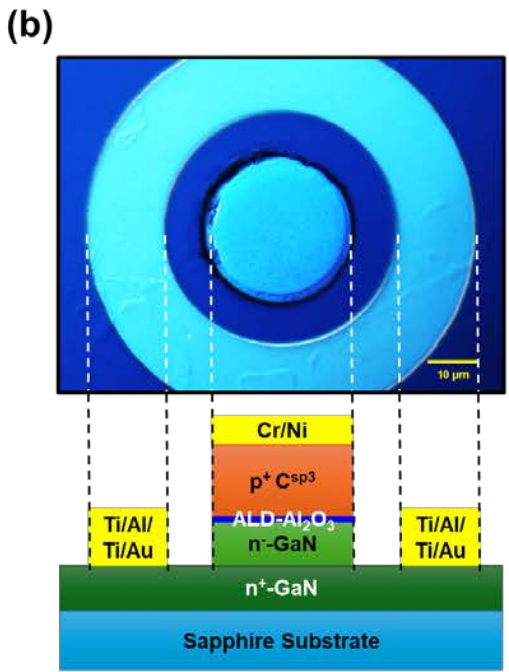

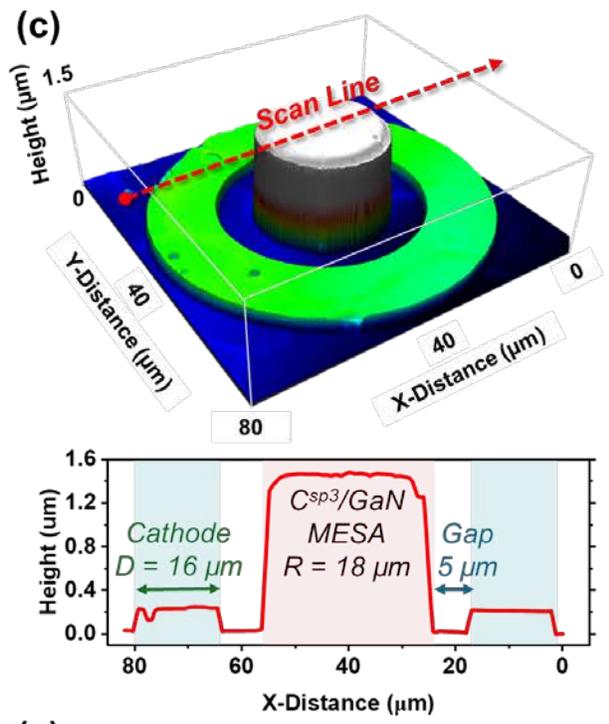

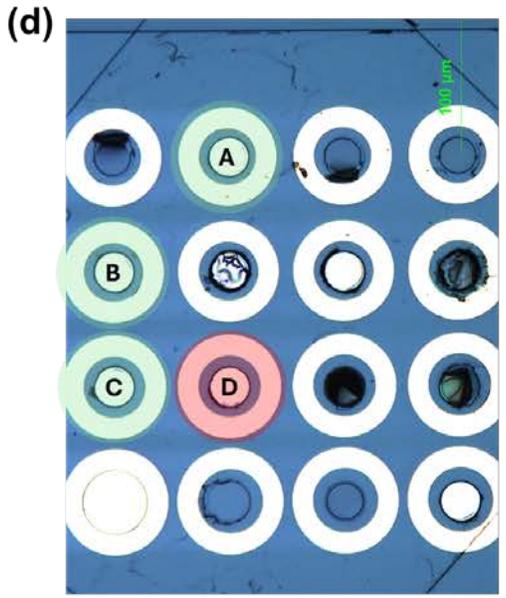

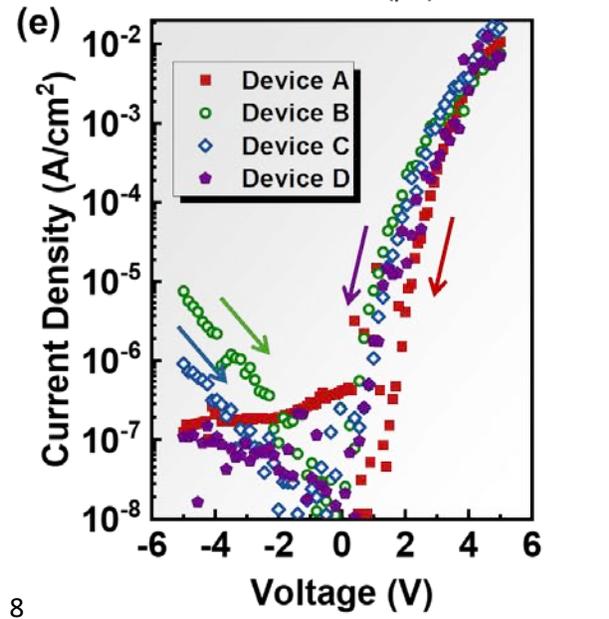

**Figure 2. Fabrication and characterization of $C^{sp3}$/GaN *pn* diodes.** (a) Schematic of the fabrication process: (i) starting from the $C^{sp3}$/GaN heterojunction, (ii) Cr/Ni deposition as *p*-contact and etch mask, (iii) two-step dry etching to form mesa and expose $n^+$-GaN, and (iv) Ti/Al/Ti/Au deposition for *n*-contact. (b) High-magnification differential interference contrast (DIC) image of a fabricated diode (Device D), with schematic cross-section below. (c) 3D optical profile of the diode measured by a Filmetrics® profiler. (d) Optical micrograph of the diode array, showing four devices (A–D) for electrical measurements and one device (D) reserved for STEM analysis. (e) *J–V* characteristics of Devices A–D. The arrows indicate the voltage sweeping directions.

Circular *p–n* diodes were fabricated from the $C^{sp3}$/GaN heterostructure, as schematically shown in **Fig. 2(a)**. Owing to the limited size of the transferred $p^+$-diamond membrane, only four diodes were realized. Beginning with the $C^{sp3}$/GaN stack [**Fig. 2(a i)**], a Cr/Ni (20/300 nm) bilayer was sputtered onto the diamond to serve as both the *p*-contact and a hard mask for subsequent etching [**Fig. 2(a ii)**]. Although Ti/Au or Ti/Pt/Au stacks generally provide superior ohmic contact to *p*-type diamond[46], Cr/Ni was chosen here because it offers more robust etch resistance[47,48] for the self-aligned dry etching step.

Mesa isolation was achieved via a two-step dry etch: $O_2/CF_4$ plasma to remove exposed $p^+$-diamond[47,48], followed by $BCl_3/Cl_2/Ar$ plasma to etch the $n^-$-GaN layer (250 nm, Si-doped at $9 \times 10^{17}$ cm$^{-3}$) and expose the $n^+$-GaN contact layer (1500 nm, Si-doped at $1.5 \times 10^{19}$ cm$^{-3}$) [**Fig. 2(a iii)**][32]. Finally, Ti/Al/Ti/Au (20/100/45/100 nm) was deposited by e-beam evaporation and patterned by liftoff to form the n-contact [**Fig. 2(a iv)**].

Device morphology confirmed the intended structure: differential interference contrast (DIC) microscopy revealed the circular mesa layout [**Fig. 2(b)**], and optical profilometry faithfully reproduced the designed height profile [**Fig. 2(c)**].

The electrical performance of four fabricated $C^{sp3}$/GaN diodes [**Fig. 2(d)**] was evaluated using a Keithley 4200 Semiconductor Parameter Analyzer. The *J–V* characteristics are shown in **Fig. 2(e)**, with key parameters summarized in **Table 1**. The devices exhibit ideality factors around 1.55 (Device B), a rectification ratio of $8.56 \times 10^4$ (Device A), and turn-on voltages between 3.8 V and 4.4 V (Supplementary **Fig. S3**), determined by extrapolating the linear region of the forward *J–V* curves. Reverse leakage current densities are in the $10^{-6}$ to $10^{-7}$ A/cm² range at −5 V. The ON-resistance is ~$10^2$ Ω·cm² at +5 V.

**Table 1.** Summary of the device parameters measured from the $C^{sp3}$/GaN *pn* diodes.

| Device | Ideality Factor ($n$) | $J_{ON}/J_{OFF}$ Ratio (±5 V) | Turn-on Voltage (V) | OFF-Current Density (A/cm²) | ON-Current Density (A/cm²) | ON-Resistance (Ω·cm²) | $D_{it}$ (cm⁻²·eV⁻¹) (0V) |
|---|---|---|---|---|---|---|---|
| A | 1.03* | $8.56 \times 10^4$ | 3.80 | $1.25 \times 10^{-7}$ | $1.07 \times 10^{-2}$ | $3.54 \times 10^2$ | $3.14 \times 10^{11}$ |
| B | 1.55 | $1.00 \times 10^3$ | 3.78 | $7.47 \times 10^{-6}$ | $7.49 \times 10^{-3}$ | $5.05 \times 10^2$ | $4.73 \times 10^{11}$ |
| C | 1.17* | $1.73 \times 10^4$ | 4.20 | $9.11 \times 10^{-7}$ | $1.58 \times 10^{-2}$ | $2.66 \times 10^2$ | $5.93 \times 10^{11}$ |
| D | 1.29* | $6.48 \times 10^4$ | 4.38 | $1.08 \times 10^{-7}$ | $6.99 \times 10^{-2}$ | $6.27 \times 10^2$ | N.A. |

*\* These extracted ideality factors may not be accurate due to the scattering of the data points within the ideal region.*

This resistance, significantly higher than in previously reported grafted diodes (Si/Ga₂O₃: 65.7 mΩ·cm² [29]; GaAs/Ga₂O₃: 2 mΩ·cm² [23]), is attributed primarily to the high contact resistance of Cr/Ni on diamond (not directly measurable here due to the limited membrane size, see Supplementary **Fig. S2**) and to interfacial element diffusion (see below). Additionally, no post-metallization annealing was applied to avoid compromising the diamond/GaN interface. These factors collectively contribute to the high contact resistance, elevated ideality factor, and relatively low forward current density (~$10^{-2}$ A/cm²).

The reverse-bias *J–V* characteristics also show hysteresis behavior under different sweeping directions (also see Supplementary **Fig. S4**), which is speculated to be caused by the insufficient passivation of the interface traps. However, under the same voltage sweeping direction, the *J-V* characteristics are consistent (see Supplementary **Figs. S5** and **S6**).

The capacitance–voltage (*C–V*) characteristics of the three diodes (Device D was used for STEM before *C–V* measurements) were measured at frequencies from 10 kHz to 2 MHz, from which the interface trap density ($D_{it}$) values were extracted (Supplementary **Fig. S7** and **Note S1**).[49] Device parameters, including $D_{it}$, are summarized in **Table 1**. Although ALD-Al$_2$O$_3$ is an effective passivation dielectric for both diamond[50,51] and GaN[28,32], its effectiveness is highly dependent on the applied thermal budget. In this study, the prolonged annealing step (350 °C for 3 h) required for interfacial bonding likely promoted Al, O and N diffusion, thereby reducing the passivation efficiency of the ALD-Al$_2$O$_3$ layer.

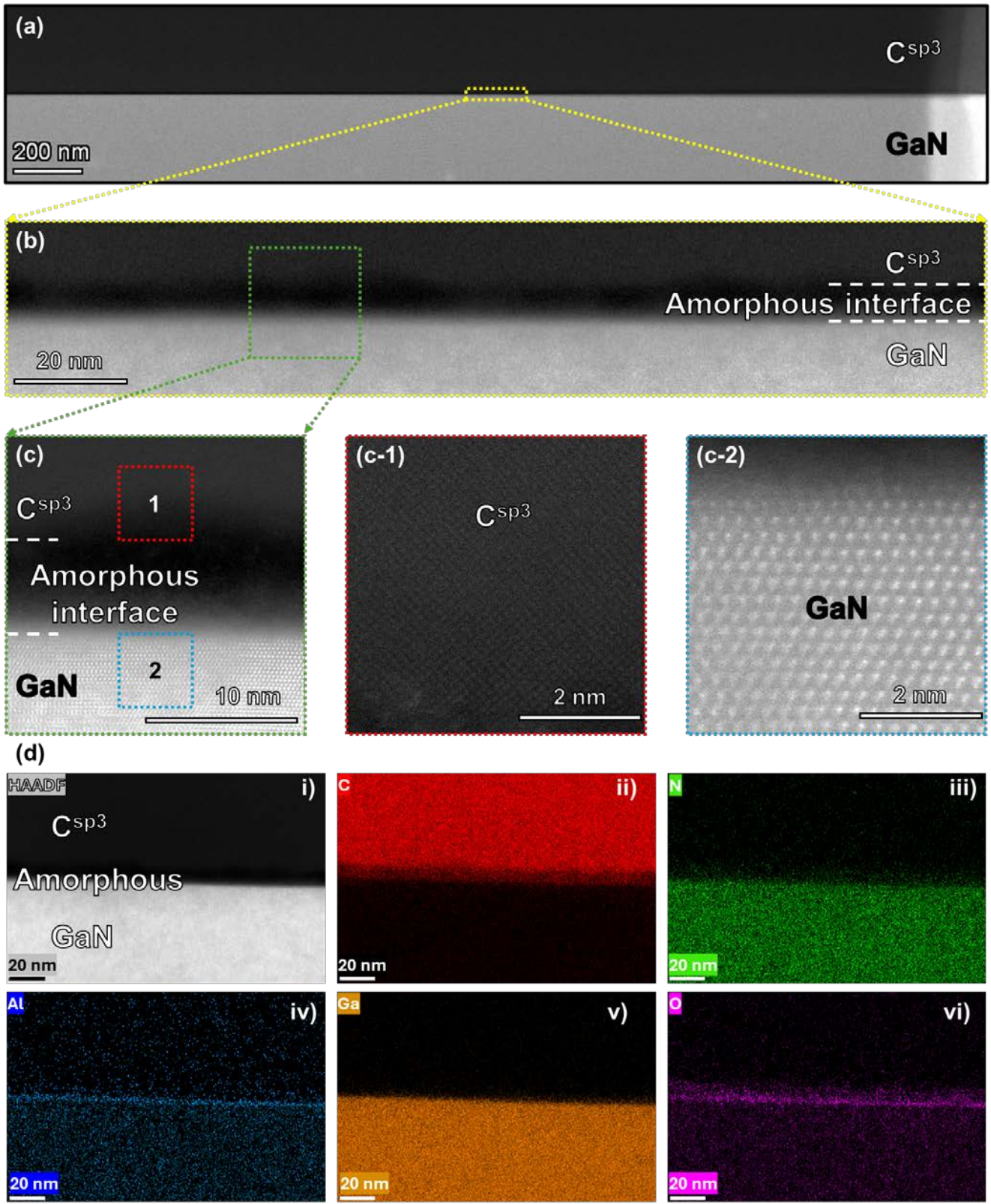

**Figure 3. HAADF-STEM and EDS analysis of the grafted $C^{sp3}$/GaN heterointerface.** (a) Low-magnification HAADF-STEM image showing interface uniformity over a micrometer-scale region. (b) Magnified view of the interface area highlighted in (a). (c) High-resolution image of the region in (b), with (c-1) and (c-2) showing atomic-resolution views of the diamond and GaN lattices, respectively. (d) Elemental mapping across the $C^{sp3}$/GaN heterointerface: (i) HAADF-STEM image, and (ii–vi) distributions of C, N, Al, Ga, and O, respectively.

To assess this effect, high-resolution HAADF-STEM was performed on Device D [see also **Fig. 2(d)** and Supplementary **Fig. S8**], which exhibited *J–V* curves nearly identical to Device A (Supplementary **Fig. S6**). A low-magnification micrograph [**Fig. 2(a)** and Supplementary **Fig. S9**] shows the separated diamond NM with a thickness of 860 nm, corresponding to the top damage-free diamond layer thickness (**Fig. S1b**). Large-area inspection revealed no observable voids or defects. A magnified view [**Fig. 2(b)**] and atomic-resolution image [**Fig. 2(c)**] confirmed a continuous heterointerface, with an amorphous interlayer measuring ~6.2 nm. The as-deposited ALD $Al_2O_3$ interlayer is 0.5 nm thick, but the interlayer thickness increases to ~6.2 nm after annealing due to chemical bonding and elemental diffusion, exceeding the 2–2.5 nm typically observed in our previous studies[23,29–31,33]. Slight misalignment between the diamond and GaN lattices was also observed, but precise crystallographic alignment is not required for grafting, as the two lattices are separated by the interlayer. Nevertheless, the cubic structure of diamond [**Fig. 2(c-1)**] and the wurtzite structure of GaN [**Fig. 2(c-2)**] remain clearly resolved, confirming preserved crystallinity in both materials.

STEM EDS (X-ray energy dispersive spectroscopy) spectrum imaging (SI) further mapped the distribution of C, N, Al, Ga and O across the heterointerface [**Fig. 2(d),** Supplementary **Fig. S10**], revealing a distinct interface along with apparent diffusion of nitrogen (N) [**Fig. 2(d iii)**], aluminum (Al) [**Fig. 2(d iv)**] and oxygen (O) [**Fig. 2(d vi)**]. Such extensive diffusion was absent in prior grafted heterojunctions[29,33,35,36] and is attributed to the extended annealing time (350 °C, 3 h) compared with earlier studies (350 °C, 5 min[28,32]). The higher thermal budget was intentionally applied here to ensure robust bonding, given the greater rigidity of diamond membranes[9] relative to Si[32] and GaAs[23].

This elemental diffusion likely degraded surface passivation, leaving a relatively high density of interface traps and contributing to the elevated interface resistance, non-ideal $J$–$V$ behavior, and relatively high $D_{it}$ values observed in the diodes, which is manifested in the hysteresis observed in the reverse-bias current $J$-$V$ sweeping direction of Device D (Supplementary **Fig. S4**). In addition, the possible degradation of diamond electronic properties (e.g., compensation of boron doping) due to the nitrogen diffusion is known at present. Future optimization should focus on identifying a thermal budget that balances strong chemical bonding with minimal diffusion. Despite these limitations, this work provides a demonstration of $C^{sp3}$/GaN heterojunction diodes formed by grafting, underscoring the feasibility of the approach and highlighting clear pathways for performance improvement.

In conclusion, we demonstrate the preliminary fabrication of a diamond–GaN heterojunction via semiconductor grafting. Circular *pn* diodes exhibit an ideality factor of 1.55, a rectification ratio of ~$10^4$, a turn-on voltage of ~4 V, and a leakage current density as low as $10^{-7}$ A/cm$^2$. AFM, XRD, and Raman measurements confirm that the single crystallinity of both diamond and GaN is preserved, while STEM reveals possible incomplete interface passivation due to elemental diffusion during annealing. Although device performance is currently limited by contact resistance and thermal processing, further optimization of grafting conditions and metallization is expected to enable high-performance diamond–GaN devices.

**ACKNOWLEDGEMENTS**

The work was supported by the Defense Advanced Research Projects Agency (DARPA) under contract 140D04-24-C-0061. The views, opinions and/or findings expressed are those of the authors and should not be interpreted as representing the official views or policies of DARPA or the U.S. Government. The STEM data were collected by using a Thermo-Fisher probe corrected Spectra 300 in STEM mode at 300 keV, which was supported by the University of Michigan and is operated by the Michigan Center for Materials Characterization.

**DATA AVAILABILITY**

The data that supports the findings of this study are available from the corresponding author upon reasonable request.